\documentclass[12pt]{article}
\usepackage{graphicx}
\usepackage{rotating}
\begin{document}
\thispagestyle{empty}
\begin{flushright}
\texttt{hep-ph/0503010} \\
\texttt{CU-PHYSICS-05/2005}\\
\end{flushright}
\vskip 50pt
\parindent 0pt

\begin{center}
{\large {\bf Matter effects on Majorana neutrino phases}}

\vskip 10pt
\renewcommand{\thefootnote}{\fnsymbol{footnote}}

{\sf Amitava Raychaudhuri$^1$ and Shashank M. Shalgar$^2$\footnote{E-mail address: shashank@bose.res.in}}

\vskip 8pt

{\em $^1$Department of Physics, University of Calcutta,\\ 
92 Acharya Prafulla Chandra Road, Kolkata 700 009, India }

\vskip 8pt

{\em $^2$S. N. Bose National Centre For Basic Sciences,\\ 
Salt Lake City, Kolkata 700 098, India}

\vskip 60pt

{\bf ABSTRACT}

\end{center}

We consider the effect of ambient matter on the Majorana phase of
neutrinos.  We find that this can lead to an observable signal if
a neutrino oscillation experiment could be performed where the
source and the detector are at appropriately different matter
densities.  We illustrate the situation using a beta beam
neutrino source as an example and show that a 5$\sigma$ signal
for the matter modification of the Majorana phase could be
possible in a 5-year run.

\vskip 30pt

\begin{center} 
\end{center}

\newpage

\renewcommand{\thesection}{\Roman{section}}
\setcounter{footnote}{0}
\renewcommand{\thefootnote}{\arabic{footnote}}

\section{Introduction}

CP non-conservation continues to remain an intriguing aspect of
particle physics.  Since its original observation in the neutral
$K$ meson system, further progress has been made more recently
through the establishment of direct CP-violation and also its
verification in $B$-meson decays. CP-violation is incorporated in
the Standard Model through the Cabibbo, Kobayashi,
Maskawa (CKM) \cite{ckm} mixing among three generations of massive
quarks involving one phase angle.  

There is no experimental signature of CP-violation in the
leptonic sector.  The extension of the CKM concept to leptons was
precluded by the long-standing belief that neutrinos are
massless, permitting mixing to be rotated away.   This view has
changed in the past decade -- results from the atmospheric,
solar, reactor, and accelerator neutrino experiments give strong
evidence for non-vanishing masses \cite{nmass}. In this
situation, it is but natural to expect CP-violation to show
up in leptons as well.  This has attained especial importance
because the large-scale neutrino experiments being now planned
for the future will be geared to permit precision measurements of
the neutrino mass and mixing parameters, including CP-violation.

In this work we examine CP phases in the neutrino sector,
focusing on two aspects: the additional `Majorana phases' and the
matter induced Mikheev-Smirnov-Wolfenstein (MSW)
\cite{wolf}, \cite{mikh} contribution to these.  We
show that the matter effects may play an important role in
revealing these phases, which are characteristic of Majorana
neutrinos.

\section{Majorana phases}

Neutrinos are electrically neutral, so they can be
self-conjugate. Such a   Majorana neutrino \cite{majorana} is
also absolutely devoid of other charges like lepton number as no
quantum number could serve the purpose of differentiating a
neutrino from an anti-neutrino.  There are several ongoing
experiments  searching for neutrinoless double beta decay events.
Incontrovertible positive evidence here will lend support to the
Majorana nature of neutrinos \cite{double}.

Majorana neutrinos offer a richer prospect for CP-violation as
more phases are permitted for the same number of generations
\cite{doi}, \cite{Kayser:1984ge}, \cite{Kayser:1983wm}. For
example, CP-violation becomes possible even for two lepton
generations.  These phases have the following origin.

Lepton number is a conserved Noether charge arising from  the
continuous symmetry: 
\begin{equation}
\psi_{i} \rightarrow e^{i\alpha_{i}}\psi_{i} \Rightarrow
\mathcal{L} \rightarrow \mathcal{L}.\label{eq:lag}
\end{equation}
The index $i$ runs over the generations and $\psi$s are the
lepton fields.  The transformation in Eq. (\ref{eq:lag}) permits
the removal of some phases from the leptonic mixing matrix, as is done
for quarks in arriving at the CKM matrix. Majorana
nature of the neutrinos would  imply the absence of this
symmetry, thereby introducing additional observable phases, apart
from the usual neutrino masses and mixing angles. For $n$
generations of neutrinos, there will be $n-1$ such additional
observables, since only phase differences have physical
significance \cite{Bernabeu:1982vi}. In principle, these can be
measured in neutrino experiments. In this paper we consider 
two generations, though the basic argument can be readily
extended.

The additional observable phases,  often termed Majorana
phases, are CP odd, that is, they change sign under CP
transformation. The phase differences show up in 
CP asymmetries of processes which, in the
simplest cases, involve at least two different amplitudes. These
amplitudes each carry a CP even and a CP odd phase and the
difference between the two CP even (CP odd) phases must be
non-zero, e.g.,
\begin{equation}
A = A_{1}e^{i(\gamma_{1}+\beta_{1})}+A_{2}e^{i(\gamma_{2}+\beta_{2})},
\label{eq:phase}
\end{equation}
where $\gamma_{i}$ are CP even and $\beta_{i}$ are CP odd
($i$=1,2) and further $\gamma_{1} \neq \gamma_{2}$, 
$\beta_{1} \neq \beta_{2}$. 

Neutrinoless double beta decay and neutrino-anti-neutrino
oscillations are examples of processes where these Majorana
phases appear \cite{Bernabeu:1982vi}, \cite{Schechter:1981bd}. 
Of these, the former is being looked for in
several experiments \cite{double}. However, it has been noted
\cite{barger}, \cite{Pascoli:2002qm}, \cite{Pascoli:2001by},
\cite{Rodejohann:2002ng} that the present constraints on neutrino
masses and mixings make it unlikely for the Majorana phases to be
determined through this route.  A process where the Majorana
phases can be measured, possibly only in principle, is the lepton
number violating phenomenon loosely termed as neutrino
anti-neutrino oscillation \cite{Bernabeu:1982vi}. Here, a massive
left-handed neutrino, emitted in a beta decay along with a
charged anti-lepton, undergoes a chirality flip and is detected in an
inverse beta decay.  In this process a charged anti-lepton is emitted
during detection, as against a charged lepton in neutrino
oscillations. The probability of such a process, symbolically
denoted by $P({\nu}_{e}\rightarrow
\bar{\nu}_{\mu})$, is \cite{degouvea}
\begin{eqnarray}
P({\nu}_{e}\rightarrow \bar{\nu}_{\mu}) &=&\left(\frac{\sin
2\theta}{2E}\right)^{2}
\left[m_{1}^{2}+m_{2}^{2}-2m_{1}m_{2}\cos\left(\frac{\Delta
m^{2}_{12}}{2E}t-\alpha\right)\right],\label{eq:anumu}
\end{eqnarray}
where, the neutrino flavour eigenstates are indicated by
$\nu_{e}$ and $\nu_{\mu}$ for simplicity of notation, though the
equation is valid for any two generations.  The energy of the
neutrinos is $E$, the mass squared difference $\Delta m^{2}_{12}
= m_{2}^{2}-m_{1}^{2}$, and $\theta $ is the mixing angle.
$\alpha$ ($\equiv (\beta_1 - \beta_2)$ of Eq.
(\ref{eq:phase})) is the CP-violating observable Majorana phase
difference\footnote{$(\frac{\Delta m^{2}_{12}}{2E}t)  \equiv
(\gamma_1 - \gamma_2)$ of Eq. (\ref{eq:phase}).}.   Eq.
(\ref{eq:anumu}) gives the probability of a chirality flip
together with flavour oscillation, given the mass eigenvalues
$m_{i}$ and the mixing angle $\theta$. A similar equation can be
written for the probability of chirality flip alone. Neither of
these phenomena is truly viable from the point of
view of present-day experiments due to the  helicity suppression
factors, $\frac{m_{i}}{E}$, in Eq. (\ref{eq:anumu}).

It should be noted that  in the two generation case, if either
one of the mass eigenvalues is vanishing then the Majorana phase
can be removed.  Unless stated otherwise, the mass eigenvalues
are assumed to be non-zero.\label{sec:maj}

\section{The MSW mechanism and Majorana Phases}

The Majorana phase cannot induce any CP asymmetry in vacuum
neutrino oscillations. To explore the situation in the presence
of ambient matter, we first summarize the MSW mechanism in some
detail, keeping in mind the Majorana property of the neutrinos.

In terms of a conventional neutrino state $\nu$ and its charge
conjugate $\nu^c$, a Majorana neutrino  is\footnote{More
generally, $N \equiv (e^{i\phi}\nu \pm e^{-i\phi}\nu^c)/\sqrt{2}$.}
$N \equiv (\nu \pm
\nu^c)/\sqrt{2}$. One of these combinations is assumed to be
light and is the focus of our attention. In see-saw models of
neutrino mass, the other is heavy and we do not consider it any
further. In discussing the matter interactions of such a Majorana
neutrino, $N$, one has to bear in mind that its left chiral
projection is a superposition of a member of an $SU(2)_L$ doublet
($\nu_L$) and a sterile state ($\nu^c_L$). Below we refer to $N_L$ as
the {\em neutrino} ($\nu$) to indicate the active component. By the
same token, in $N_R$ only the $\nu^c_R$ participates in weak
interactions. We refer to it as the {\em anti-neutrino}
($\bar\nu$) in the following.

Neutral current interactions are the same for all
active neutrinos and do not affect neutrino oscillations. The
effective propagator of electron neutrinos is modified due
to charge current interactions with the ambient matter. Ignoring
mixing for the moment, the matter-modified
energy for spin $\frac{1}{2}$ electron neutrinos can be written
as\footnote{On the RHS of the arrow are the matter-modified
quantities hereafter.},
\begin{equation}
E=\mathbf{\alpha}.\mathbf{p}+\beta m \rightarrow 
E=\mathbf{\alpha}.\mathbf{p}+\beta m+\sqrt{2}G_{F}n_{e},
\label{en:mod}
\end{equation}
which follows from \cite{pal}, \cite{notzold}, \cite{nieves}, 
\begin{equation}
G(p)=\frac{i}{\not p-m} \rightarrow 
\frac{i}{\not{p}-m-\gamma^{0}\sqrt{2}G_{F}n_{e}}\;\;.
\label{one:pro}
\end{equation}
The above result, obtained using finite temperature field theory
methods, is valid in the rest frame of the ambient matter.
For a Majorana neutrino, the propagator
contains additional terms involving chirality flip. However,
these terms will not be important for the subsequent discussion.

For Majorana neutrinos, the particle cannot be
distinguished from the antiparticle. CP-violation due to the
Majorana phase $\alpha$ is a characteristic of the spin-half nature
of the neutrinos. This is evident from the vital role played by
chirality flip in the CP-violating neutrino-anti-neutrino
oscillation phenomenon.  This is unlike CP-violation due to a
Dirac phase, which, in principle, could also be formulated for
spin zero particles.

The effect of mixing can be incorporated by promoting $m$ to a
$(2 \times 2)$ matrix $M_{\nu}$.  We choose to work in a basis in
which the charged lepton mass matrix is diagonal and where the
Majorana phase is included as a multiplicative factor associated
with the neutrino fields. Due to the Majorana property, these
phases cannot be absorbed into the neutrino fields and will have
observable consequences. In this basis, the neutrino mass matrix,
$M_\nu$, and the mixing matrix, $U$, are real.

In the highly relativistic limit, the effect of ambient matter on
the neutrino mass matrix can be written as
\cite{wolf}, \cite{mikh},

\begin{equation}
{M_{\nu}M_{\nu}^{\dagger}} \rightarrow 
{M_{\nu}M_{\nu}^{\dagger}}+\left(\matrix{2\sqrt{2}G_{F}n_{e}E & 0 
\cr 0 & 0}\right),
\end{equation}
in the flavour basis. 
The second term  on the RHS of the arrow -- the matter
modification -- takes a negative sign for anti-neutrinos.  The
eigenvalues of the RHS matrix  are the squared mass eigenvalues
in the medium, whereas the two eigenvectors determine the
diagonalizing matrix and hence, the matter-modified mixing angle
$\theta_{m}$. The diagonalizing matrix is the same for Dirac as
well as Majorana neutrinos \cite{Langacker:1986jv}.  It should be
noted that this matrix is not unique. There is freedom of  adding
a phase common to a row or a column or both.  The most general
matter-modified mixing matrix can be parametrized as,
\begin{equation}
U_{m}=\left(\matrix{\cos\theta_{m}e^{i\eta_{1}} & 
\sin\theta_{m}e^{i\eta_{1}+i\eta_{2}} \cr 
-\sin\theta_{m} & \cos\theta_{m} e^{i\eta_{2}}}\right).
\end{equation}
The phases $\eta_{1}$ and $\eta_{2}$ are undetermined from the
mass matrix and cannot be fixed from the physics under
consideration. To simplify matters, it would not be unreasonable
to choose them to be independent of the mixing angles
$\theta_{m}$. It then follows that both $\eta_i$ have to be zero
in order to ensure that in the limit of ambient matter density
going to zero, the vacuum mixing matrix, $U$, is reproduced, where
\begin{equation}
U=\left(\matrix{\cos{\theta} & \sin{\theta} \cr -\sin\theta & 
\cos\theta}\right).
\end{equation}
The matter-modified eigenstates of the neutrinos can be expressed
in terms of the vacuum mass eigenstates as
\begin{equation}
\left(\matrix{e^{i\alpha_{1}}\nu_{1m} \cr 
e^{i\alpha_{2}}\nu_{2m}}\right)=
U_{m}^{\dagger}\left(\matrix{\nu_{e} \cr \nu_{\mu}}\right)=
U_{m}^{\dagger}U\left(\matrix{\nu_{1} \cr e^{i\alpha}\nu_{2}}\right)
=\left(\matrix{\cos\phi & \sin\phi 
\cr -\sin\phi & \cos\phi}\right)\left(\matrix{\nu_{1} 
\cr e^{i\alpha}\nu_{2}}\right).
\end{equation}
where $\phi=\theta_{m}-\theta$.

The phases can be readily found to be
\begin{eqnarray}
\tan\alpha_{1}=\frac{\sin{\phi}\sin{\alpha}}{\cos\phi+\sin\phi\cos\alpha},
\\
\tan\alpha_{2}=\frac{\cos\phi\sin\alpha}{\cos\phi\cos\alpha-\sin\phi}.
\end{eqnarray}

It should be noted that in matter the relative Majorana
phase, responsible for CP-violation,  is:
\begin{eqnarray}
\alpha^{\prime}&=&\alpha_{2}-\alpha_{1} \nonumber \\
&=&\tan^{-1}\left(\frac{\cos\phi\sin\alpha}{\cos\phi\cos\alpha-
\sin\phi}\right)-\tan^{-1}\left(\frac{\sin{\phi}\sin{\alpha}} 
{\cos\phi+\sin\phi\cos\alpha}\right).
\label{eq:alprim}
\end{eqnarray}
It vanishes when the vacuum Majorana phase $\alpha = 0$.
Recall that $\phi=\theta_{m}-\theta$ and therefore $\alpha^\prime$ is
significantly different from $\alpha$ near a resonance.

\section{CP asymmetries in neutrino oscillations}

A consequence of the propagation eigenstates not being the same
as flavour eigenstates is the phenomenon of neutrino
oscillations. Oscillation of neutrino flavours requires a 
nontrivial mixing angle $\theta$ and non-degenerate mass
eigenvalues of the neutrino mass matrix. The probability
amplitude of neutrino oscillations,
\begin{eqnarray}
A(\nu_{e} \rightarrow \nu_{\mu}) 
= \sum_{i}\langle \mu^{-}|\gamma^{\mu}(1-\gamma_{5})U_{\mu
i}|\nu_{i}(t)\rangle
\langle \nu_{i}(0)|U_{e i}^{\ast}\gamma_{\mu}(1-\gamma_{5})|e^{-}\rangle
\end{eqnarray}
is modified due to the presence of ambient matter.
Matter effects entail the following changes for
Majorana neutrinos:
\begin{eqnarray}
\Delta m_{12}^{2} &\rightarrow& 
{\sqrt{(\Delta m_{12}^{2}\cos\theta-2\sqrt{2}G_{F}n_{e}E)^{2}+
(\Delta m_{12}^{2}\sin\theta)^{2}}}\;\;,\\
\tan 2 \theta_{m} &\rightarrow& \frac{\sin 2 \theta}
{\cos 2 \theta - 2\sqrt{2}G_{F}n_{e}E}\;\;,\\
\alpha &\rightarrow& \alpha^{\prime}.
\label{maj:mod}
\end{eqnarray}
Apart from the usual modifications made for Dirac neutrinos,
there is a further replacement given by Eq.
(\ref{maj:mod}), where $\alpha^{\prime}$ is given in Eq.
(\ref{eq:alprim}). 

In the adiabatic limit, the probability of neutrino oscillations
in matter for the two-flavour case is given by
\begin{equation}
P(\nu_{e},0 \rightarrow \nu_{\mu},t) = 
\left|U_{e1}^{\ast}(0)U_{\mu 1}(t)e^{i\int
E_{1}(t)dt-i\alpha_{1}}+ U_{e2}^{\ast}(0)U_{\mu 2}(t)e^{i\int
E_{2}(t)dt+i\alpha-i\alpha_{2}}\right|^{2}.
\end{equation}
The dependence of the effective energy and the mixing angle on
time is a result of the varying matter density.  It should be
noted that the Majorana phase, like the mixing matrix, is not a
kinematical quantity. It plays a role only at emission and
detection and hence, unlike the effective energy, is not
integrated over time.

Assuming for simplicity that 
the point of emission is in vacuum,  the oscillation probability is
\begin{eqnarray}
P&=&\cos^{2}\theta\sin^{2}\theta_{m}+\sin^{2}\theta\cos^{2}\theta_{m}
\nonumber \\
&-&2\frac{\sin 2\theta}{2}\frac{\sin 2\theta_{m}}{2}\cos\left(\int_{0}^{t} 
\frac{m_{12m}^{2}}{2E}dt+(\alpha-\alpha^{\prime})\right).
\end{eqnarray}
It is seen that the dependence on $(\alpha-\alpha^{\prime})$
vanishes if the point of detection is also in vacuum. The same
result holds even in the case of identical matter densities at
the point of emission and  detection, including the special case
of uniform matter density. Different densities at the source and
detection points is therefore a must for the Majorana phase to be
effective in this process.

Using the CP odd nature of $\alpha$, we arrive
at an expression for the CP asymmetry:
\begin{equation}
\Delta P = |P - \overline{P}| = {\sin 2\theta}{\sin
2\theta_{m}}\sin\left(\int_{0}^{t}
\frac{m_{12m}^{2}}{2E}dt\right)\sin(\alpha-\alpha^{\prime}).
\label{cp:asym}
\end{equation}
Here $\overline{P}$ is the oscillation probability of anti-neutrinos.

\section{Numerical estimates}

Phenomenologically, it is the CP asymmetry of a process that can
give evidence of CP-violation in a system. However, in the case of
neutrinos propagating through matter, the asymmetry,
\begin{equation}
\Delta P = P(\nu_{e},0;\nu_{\mu},L)-P(\bar{\nu}_{e},0;\bar{\nu}_{\mu},L),
\end{equation}
gets contribution from CP-violation as well as the overwhelming
dominance of ambient matter over antimatter.
This implies that the CP asymmetry given by Eq. (\ref{cp:asym})
will need a modification to accommodate the fact that the value of
$\theta_{m}$ will be different for 
neutrinos and anti-neutrinos. Hence, the study of CP-violation
through the CP asymmetry would require a disentangling of the
contribution due to matter effects and is not the most
favourable way of analytically probing the physics we are after.

Therefore, we turn to the following alternative.  As noted
earlier,  $\alpha^{\prime} - \alpha$  is appreciable only at
certain energies -- near a resonance -- and that too for either
neutrinos or anti-neutrinos depending on whether the mass-squared
splitting is positive or negative.  Given a density profile, for
appropriate energies the oscillation probability will then
deviate from the one estimated without taking the
Majorana phases into account (i.e., $ \alpha$ = 0). To
experimentally look for this phase it will be useful to have a
beam of neutrinos of well-defined energy spectrum. As an
illustration, we consider neutrinos from a beta beam
source\footnote{A beta beam is a collimated neutrino source
produced through the decay of accelerated beta-active nuclei.}
\cite{Zucchelli:2002sa}, \cite{Albright:2004iw}.  We consider the
$\nu_e$ ($\bar\nu_e$) survival probability and examine its
sensitivity to matter effects through $\alpha$. The expectation
for the number of events for a 5-year run has been illustrated in
Fig. 1 for a beta beam source at CERN with a 440kT Water Cerenkov
detector\footnote{100\% efficiency has been taken.} at Gran
Sasso. In case of Water Cerenkov detectors  (e.g.,
Super-Kamiokande) energy resolution better than 50 MeV can be
achieved, which is assumed in our analysis.  Here $\gamma=580$
($\gamma=350$) has been chosen and $5 \times 10^{11}$ ($10^{13}$)
decays per second have been assumed for the neutrino
(anti-neutrino) beam.  The matter density is taken to vary
linearly\footnote{The densities chosen for this example are close
to the matter density of the earth but the linear variation does
not reflect the true profile.} from 2.5 gm/cc to 2.8 gm/cc. We
use the mass splittings favoured by the solar and atmospheric
neutrino experiments and find that a matter induced enhancement
of $\alpha$ is possible only for the $\nu_e$ ($\bar\nu_e$) to
$\nu_\tau$ ($\bar\nu_\tau$) transition for normal (inverted)
hierarchy. Under these circumstances, it is justified to use the
two-flavour oscillation formula, as
$P(\nu_{e}\rightarrow\nu_{\mu}) \ll
P(\nu_{e}\rightarrow\nu_{\tau})$.  The oscillation probability,
needless to say, depends on the value of $\theta$ -- which we
have chosen to be $5^{\circ}$; well within the experimental bound
for $\theta_{13}$ -- and may vanish for a particular combination
of $\theta$ and $L$.  The modification in event rate is prominent
at energies which are not enough to produce tau leptons.
However, the effect of the Majorana phase should be detectable at
more than 5$\sigma$ level (for $\alpha=45^{\circ}$) in a
disappearance experiment of electron neutrinos once the other
mixing angles and mass-splittings are measured to sufficient
precision.  Note that the deviation in flux due to $\alpha$ is
appreciable only if the matter densities at the point of emission
and detection are close to the resonance value.  For a
given vacuum mixing angle, to observe this effect at higher
energies the matter densities would have to be lower and vice
versa.  In the example displayed in Fig. 1, there is no impact of
the Majorana phase at high energies.  In fact, such an energy
dependence can be considered as an evidence for the Majorana
property of the neutrinos, though the absence of such an effect
would not demonstrate the reverse.
\begin{figure}
\begin{scriptsize}
\input{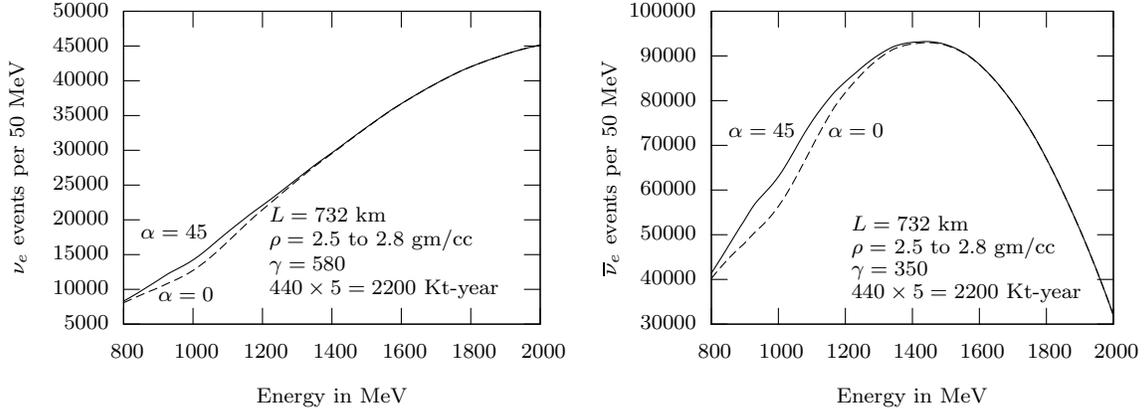}
\end{scriptsize}
\caption{\sf \small $\nu_{e}$ and $\bar{\nu}_{e}$ event rates for
Majorana phase $\alpha=45^{\circ}$ (solid) and $\alpha=0^{\circ}$
(dashed) for a 5-year beta beam run. A linear increase of the
matter density from 2.5 gm/cc to 2.8 gm/cc over 732 km is
assumed. $\Delta m^{2}=+2 \times 10^{-3}$ eV$^2$ (left
panel), $\Delta m^{2}=-2 \times 10^{-3}$ eV$^2$ (right
panel) and $\theta=5^{\circ}$ are used.}
\end{figure}

\section{Conclusions}

The possibility of observing the Majorana phase in neutrino
oscillations through a matter-induced effect may be difficult due
to the practical limitation of the availability of  a suitable
density profile.  Nonetheless, it may not be out of reach of
proposed precision experiments.  Oscillation of neutrinos in the
presence of ambient matter may thus play an important role in
revealing the Majorana nature of neutrinos.

\vskip 20pt

\parindent 0pt

{\large{\bf {Acknowledgements}}}

The work of AR has been supported in part by the Department of
Science and Technology, Government of India. SMS would like to thank
the Council for Scientific and Industrial Research, India for a
fellowship.


\vskip 20pt


\begin{thebibliography}{99}
\bibitem{ckm} M.~Kobayashi and T.~Maskawa, Prog.\ Theor.\ Phys.\
{\bf 49}, 652 (1973).
\bibitem{nmass}
%
Q.~R.~Ahmad {\it et al.}  [SNO Collaboration],
%
Phys.\ Rev.\ Lett.\  {\bf 89}, 011301 (2002)
[arXiv:nucl-ex/0204008];
%
S.~Fukuda {\it et al.}  [Super-Kamiokande Collaboration],
%
Phys.\ Lett.\ B {\bf 539}, 179 (2002)
[arXiv:hep-ex/0205075];
B.~T.~Cleveland {\it et al.},
%
Astrophys.\ J.\  {\bf 496}, 505 (1998);
W.~Hampel {\it et al.}  [GALLEX Collaboration],
Phys.\ Lett.\ B {\bf 447}, 127 (1999);
K.~Eguchi {\it et al.}  [KamLAND Collaboration],
%
Phys.\ Rev.\ Lett.\  {\bf 90}, 021802 (2003)
[arXiv:hep-ex/0212021];
Y.~Fukuda {\it et al.}  [Kamiokande Collaboration],
%
Phys.\ Lett.\ B {\bf 335}, 237 (1994)
  R.~Becker-Szendy {\it et al.},
  Nucl.\ Phys.\ Proc.\ Suppl.\  {\bf 38}, 331 (1995);
W.~W.~M.~Allison {\it et al.}  [Soudan-2 Collaboration],
%
Phys.\ Lett.\ B {\bf 449}, 137 (1999)
[arXiv:hep-ex/9901024];
M.~Ambrosio {\it et al.}  [MACRO Collaboration],
%
Phys.\ Lett.\ B {\bf 434}, 451 (1998)
[arXiv:hep-ex/9807005];
M.~H.~Ahn {\it et al.}  [K2K Collaboration],
Phys.\ Rev.\ Lett.\  {\bf 90}, 041801 (2003)
[arXiv:hep-ex/0212007];
\bibitem{wolf}
L.~Wolfenstein,
Phys.\ Rev.\ D {\bf 17}, 2369 (1978).
\bibitem{mikh}
S.~P.~Mikheev and A.~Y.~Smirnov,
Nuovo Cim.\ C {\bf 9}, 17 (1986).
\bibitem{majorana} E.~Majorana,
Nuovo Cim.\  {\bf 14}, 171 (1937).
\bibitem{double} 
H.~V.~Klapdor-Kleingrothaus, A.~Dietz, I.~V.~Krivosheina and O.~Chkvorets,
Nucl.\ Instrum.\ Meth.\ A {\bf 522}, 371 (2004)
[arXiv:hep-ph/0403018];
A.~M.~Bakalyarov, A.~Y.~Balysh, S.~T.~Belyaev, V.~I.~Lebedev and S.~V.~Zhukov
                  [C03-06-23.1 Collaboration],
arXiv:hep-ex/0309016;
C.~E.~Aalseth {\it et al.}  [16EX Collaboration],
Phys.\ Rev.\ D {\bf 65}, 092007 (2002)
[arXiv:hep-ex/0202026];
X.~Sarazin,
arXiv:hep-ex/0412012.
\bibitem{doi}
M.~Doi, T.~Kotani, H.~Nishiura, K.~Okuda and E.~Takasugi,
Phys.\ Lett.\ B {\bf 102}, 323 (1981).
\bibitem{Kayser:1984ge}
B.~Kayser,
Phys.\ Rev.\ D {\bf 30}, 1023 (1984).
\bibitem{Kayser:1983wm}
B.~Kayser and A.~S.~Goldhaber,
Phys.\ Rev.\ D {\bf 28}, 2341 (1983).
\bibitem{Bernabeu:1982vi}
J.~Bernabeu and P.~Pascual,
Nucl.\ Phys.\ B {\bf 228}, 21 (1983);
J.~Schechter and J.~W.~F.~Valle,
Phys.\ Rev.\ D {\bf 23}, 1666 (1981).
\bibitem{Schechter:1981bd}
J.~Schechter and J.~W.~F.~Valle,
Phys.\ Rev.\ D {\bf 25}, 2951 (1982).
\bibitem{barger}
V.~Barger, S.~L.~Glashow, P.~Langacker and D.~Marfatia,
Phys.\ Lett.\ B {\bf 540}, 247 (2002)
[arXiv:hep-ph/0205290].
\bibitem{Pascoli:2002qm}
S.~Pascoli, S.~T.~Petcov and W.~Rodejohann,
Phys.\ Lett.\ B {\bf 549}, 177 (2002)
[arXiv:hep-ph/0209059].
\bibitem{Pascoli:2001by}
S.~Pascoli, S.~T.~Petcov and L.~Wolfenstein,
Phys.\ Lett.\ B {\bf 524}, 319 (2002)
[arXiv:hep-ph/0110287].
\bibitem{Rodejohann:2002ng}
W.~Rodejohann,
arXiv:hep-ph/0203214.
\bibitem{degouvea}
A.~de Gouvea, B.~Kayser and R.~N.~Mohapatra,
Phys.\ Rev.\ D {\bf 67}, 053004 (2003)
[arXiv:hep-ph/0211394].
\bibitem{pal}
P.~B.~Pal and T.~N.~Pham,
Phys.\ Rev.\ D {\bf 40}, 259 (1989).
\bibitem{notzold}
D.~Notzold and G.~Raffelt,
Nucl.\ Phys.\ B {\bf 307}, 924 (1988).
\bibitem{nieves}
J.~F.~Nieves,
Phys.\ Rev.\ D {\bf 40}, 866 (1989).
\bibitem{Langacker:1986jv}
P.~Langacker, S.~T.~Petcov, G.~Steigman and S.~Toshev,
Nucl.\ Phys.\ B {\bf 282}, 589 (1987).
\bibitem{Zucchelli:2002sa}
P.~Zucchelli,
Phys.\ Lett.\ B {\bf 532}, 166 (2002).
\bibitem{Albright:2004iw}
C.~Albright {\it et al.}  [Neutrino Factory/Muon Collider Collaboration],
Part of the APS Neutrino Study.
arXiv:physics/0411123.
\end{thebibliography}
\end{document}